\documentclass[aps,prl,showpacs,noshowkeys,amsmath,amssymb,amsfonts,reprint]{revtex4-1}

\usepackage{graphicx}
\usepackage{bm}

\usepackage[colorlinks=true,linkcolor=blue,citecolor=blue,urlcolor=blue]{hyperref}
\usepackage[all]{hypcap}
\begin{document}
\title{Bleb Nucleation through Membrane Peeling}
\author{Ricard Alert}
\email{ricardaz@ecm.ub.edu}
\affiliation{Departament d'Estructura i Constituents de la Mat\`{e}ria, Universitat de Barcelona, Barcelona, Spain}
\author{Jaume Casademunt}
\email{jaume.casademunt@ub.edu}
\affiliation{Departament d'Estructura i Constituents de la Mat\`{e}ria, Universitat de Barcelona, Barcelona, Spain}
\date{\today}
\begin{abstract}
We study the nucleation of blebs, i.e. protrusions arising from a local detachment of the membrane from the cortex of a cell. Based on a simple model of elastic linkers with force-dependent kinetics, we show that bleb nucleation is governed by membrane peeling. By this mechanism, the growth or shrinkage of a detached membrane patch is completely determined by the linker kinetics, regardless of the energetic cost of the detachment. We predict the critical nucleation radius for membrane peeling and the corresponding effective energy barrier. These may be typically smaller than those predicted by classical nucleation theory, implying a much faster nucleation. We also perform simulations of a continuum stochastic model of membrane-cortex adhesion to obtain the statistics of bleb nucleation times as a function of the stress on the membrane. The determinant role of membrane peeling changes our understanding of bleb nucleation and opens new directions in the study of blebs.
\end{abstract}
\maketitle
The adhesion between the cell membrane and the cytoskeleton is crucial to many physiological processes, including apoptosis \cite{Mills1998,*Coleman2001}, cell spreading \cite{Norman2010}, cytokinesis \cite{Sedzinski2011}, and motility \cite{Charras2008,*Fackler2008,*Paluch2013}. The membrane is attached to the actomyosin cortex via a number of specific linker molecules \cite{Sheetz2001}. These linkers continuously bind and unbind, and they are under stress due to both osmotic pressure and contractile tension generated by myosin in the cortex.

Blebs are cellular protrusions that form when the cell membrane locally unbinds from the underlying actomyosin cortex. Once detached, the unbound membrane inflates due to intracellular pressure, thus acquiring the shape of a spherical cap. Typically, a new cortex starts to assemble beneath the detached membrane and retracts the bleb, thus healing the membrane to the cortex again \cite{Charras2008a}. Among many other functions, membrane blebbing is often used for motility by several cell types \cite{Blaser2006}, mainly amoebae \cite{Langridge2006,*Yoshida2006,*Maugis2010} and invasive cancer cells \cite{Sahai2003}. Therefore, a physical understanding of blebbing will also provide insights into the regulation of bleb-based motility \cite{Paluch2006,*Lammermann2009,*Diz-Munoz2010,*Bergert2012,*Paluch2013,*Tozluoglu2013,*Zatulovskiy2014,*Liu2015}.

Bleb formation can be triggered internally by actomyosin contractile stresses or externally via micropipette aspiration \cite{Merkel2000,*Rentsch2000,*Brugues2010,Sliogeryte2014}, laser ablation of the cortex \cite{Tinevez2009} or osmotic shocks \cite{Sliogeryte2014}. These experimental studies show that a minimal stress is needed to detach the membrane from the cortex. This observation has been recently rationalized within a model of mem\-brane\--cor\-tex adhesion that incorporates both active cortical tension and external suction \cite{Alert2015a}. However, the spontaneous formation of blebs is a nucleation phenomenon driven by local fluctuations, and hence a membrane detachment of a minimum size is required.

Here, we propose that the determinant mechanism for bleb nucleation is membrane peeling from the cortex, whereby the membrane sequentially unbinds from adjacent linkers, a phenomenon that has been observed in bleb formation \cite{Charras2008a}. This process is controlled by the linker kinetics and completely determines the growth or decay of a detached membrane patch, regardless of the energetic cost. Within a simple model of force-de\-pen\-dent kinetics of the linkers, we predict the bleb nucleation radius and the effective energy barrier. Typically, the critical nucleation size for membrane peeling is significantly smaller than the one predicted by the classical nucleation theory, implying a strong reduction of the nucleation time scales. Based on a formulation of first-passage time statistics for the formation of the critical nucleation patch, we study the kinetics of bleb nucleation via numerical simulations.

Our study of bleb nucleation is based on the model for mem\-brane\--cor\-tex adhesion introduced in \cite{Alert2015a}. This model considers a nearly flat membrane subject to a net outward pressure $f$ and attached to the underlying static cortex by a density of bound molecular linkers $\rho_b$, smaller than the density of available linkers $\rho_0$. These linkers (such as ERM proteins) are modeled as springs of elastic constant $k$ that are fixed on the cortex, and that attach to the membrane at a constant rate $k_{\textrm{on}}$ and detach from it at a force-dependent rate $k_{\textrm{off}}$. Then, the coupled non-linear dynamics of the membrane position $u$, which measures the stretching of the bound linkers, and the density of links $\rho_b$ is given by
\vskip-0.5cm
\begin{subequations}
\label{eq 1}
\begin{align}
\eta\frac{du}{dt}&=f-\rho_b ku \label{eq 1a}\\
\frac{d\rho_b}{dt}&=k_{\textrm{on}}\left[\rho_0-\rho_b\right]-k_{\textrm{off}}\left(u\right)\rho_b, \label{eq 1b}
\end{align}
\end{subequations}
\vskip-0.1cm
\noindent where $\eta$ is an effective viscosity per unit length, and
\begin{equation} \label{eq kinetics}
k_{\textrm{off}}\left(u\right)=k_{\textrm{off}}^0 e^{\beta ku\delta},
\end{equation}
with $\delta$ being a bond length in the nanometric scale \cite{Evans2001} and $\beta=\left(k_BT\right)^{-1}$. These equations predict a mem\-brane-cor\-tex unbinding transition above a critical pressure $f^*$ given by the solution to the implicit equation $\alpha^* e^{1+\alpha^*}=\chi^{-1}$, with $\alpha\equiv f\delta\beta/\rho_0$, and $\chi\equiv k_{\textrm{off}}^0/k_{\textrm{on}}$ \cite{Alert2015a}. In terms of the density of links, the unbinding occurs below a critical density $\rho_b^*=\alpha^*\rho_0/z^*$, with $z^*$ being the solution of $z^*=\alpha^*\left(1+\chi e^{z^*}\right)$.

\begin{figure}[tbp]
\begin{center}
\includegraphics[width=\columnwidth]{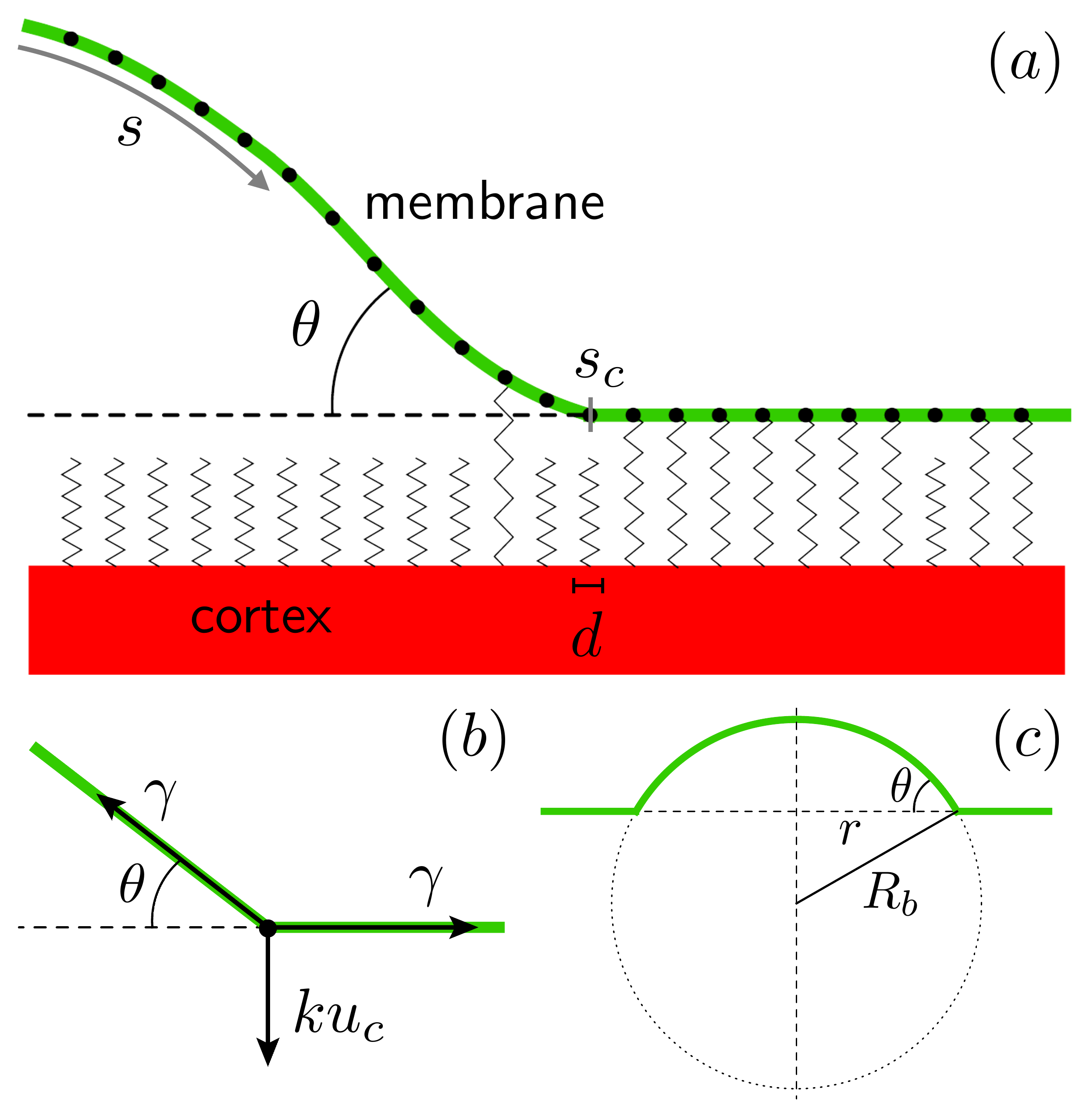}
\caption{\label{fig1}(Color online) Membrane peeling from the cortex. (a) Schematics of the adhesion between the membrane and the cortex by spring-like molecular linkers (zig-zag lines). The contact line at $s_c$ connects the unbound membrane patch (left) to the adhered membrane (right), forming a contact angle $\theta$. (b) Normal force balance at the contact line: adhesion balances the vertical pulling of the unbound membrane, Eq. \ref{eq force-balance}. (c) Geometry of the unbound membrane patch, a spherical cap of radius $R_b=2\gamma/f$ and detached radius $r=R_b\sin\theta$.}
\end{center}
\end{figure}

We now ask whether a given unbound membrane patch will grow to form a bleb or shrink. To this end we consider a detached region next to another where the membrane is attached to the cortex (Fig. \ref{fig1}a). Being $s$ the arc length coordinate along the membrane, we define the contact line as the set of points $s_c\left(t\right)$ having the critical density of links, this is $\rho_b\left(s_c,t\right)=\rho_b^*$. The speed of the contact line, $v_c=ds_c/dt$, is known as the peeling speed of the membrane. Following Dembo et al. \cite{Dembo1988} and using Eq. \ref{eq 1b} (see details in \cite{SM}), the stationary peeling speed is given by
\begin{equation} \label{eq contact-line}
-v_c\left.\frac{\partial\rho_b}{\partial s}\right|_{s=s_c}=k_{\textrm{on}}\left[\rho_0-\rho_b^*\right]-k_{\textrm{off}}\left(u_c\right)\rho_b^*,
\end{equation}
where $u_c\equiv u\left(s_c,t\right)$ is the stretching of the bound linkers at the contact line. Then, the incipient bleb will grow by peeling the membrane off the cortex if $v_c>0$, and will shrink by healing the adhesion if $v_c<0$. Therefore, since $\left.\partial\rho_b/\partial s\right|_{s=s_c}>0$ by definition (see Fig. \ref{fig1}a), peeling will occur if the stretching of the bonds at the contact line, $u_c$, exceeds the critical value $u^*=u_0z^*$, with $u_0\equiv k_BT/\left(k\delta\right)$, which solves Eq. \ref{eq contact-line} for $v_c=0$. It is important to stress that this kinematic condition is independent of the energy gain or loss associated to the motion of the contact line.

Next, without solving for the inner shape of the contact region \cite{Evans1985,*Garrivier2002}, we can establish the normal force balance condition at the contact line. Neglecting bending rigidity, the elastic force of the linkers, $ku_c$, balances the vertical pulling of the tension $\gamma$ produced by the unbound membrane at a contact angle $\theta$ (see Fig. \ref{fig1}b):
\begin{equation} \label{eq force-balance}
2\pi r\gamma\sin\theta=N_cku_c.
\end{equation}
Here, $2\pi r$ is the length of the (circular) contact line. This contains $N_c\approx 2\pi rd\rho_b^*$ bonds, with $d$ being the diameter of the effective area that a bond covers on the membrane, presumably of a few tens of nanometers (see Fig. \ref{fig1}a).

\begin{figure}[htbp]
\begin{center}
\includegraphics[width=\columnwidth]{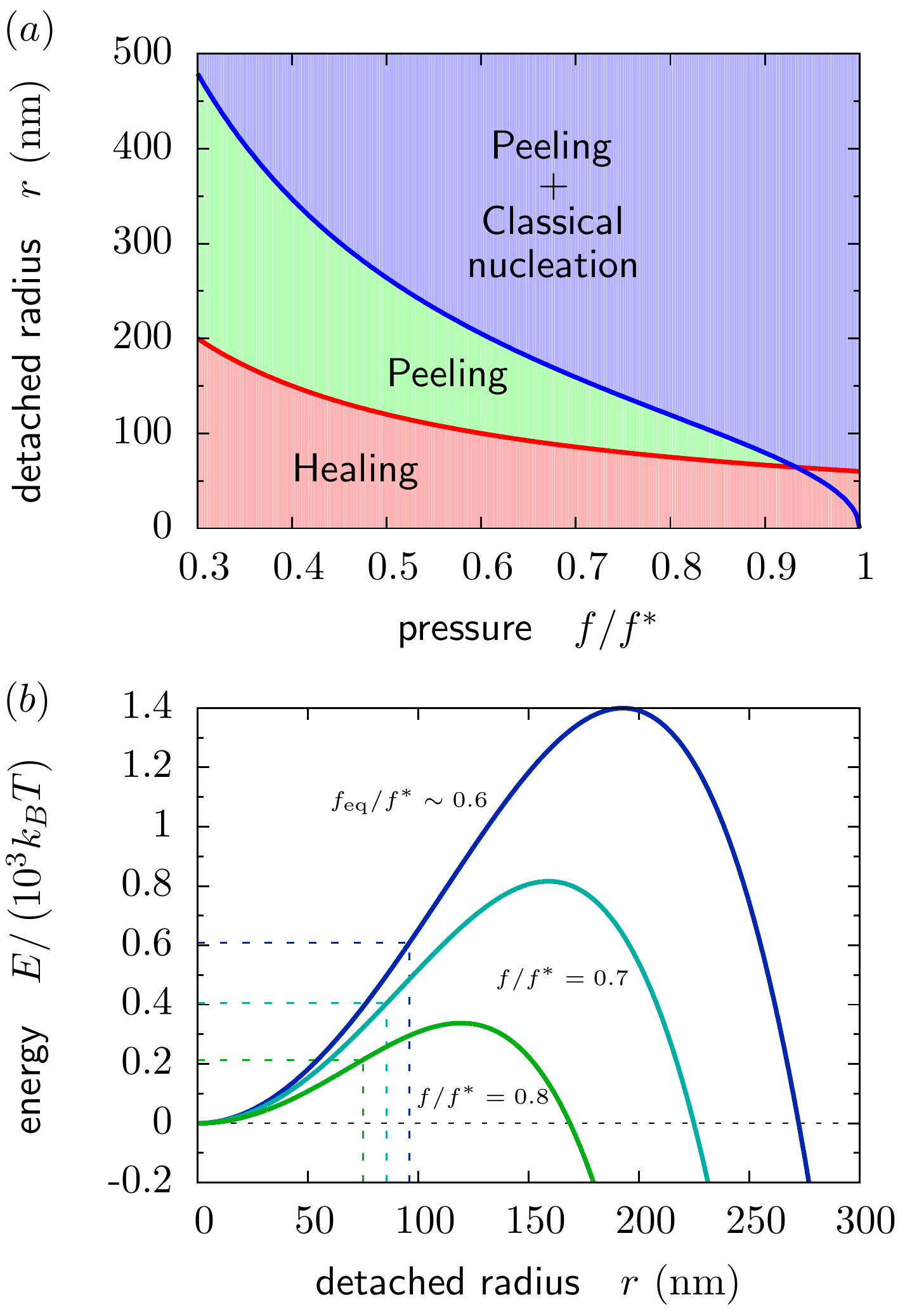}
\caption{\label{fig2}(Color online) Bleb nucleation through membrane peeling. (a) Evolution of a detached membrane patch of radius $r$ subject to a pressure $f$. Mem\-brane\--cor\-tex adhesion is healed for $r<r_p$ (red region), and the membrane is peeled from the cortex for $r>r_p$ (green and blue regions), Eq. \ref{eq critical-radius}. Classical bleb nucleation would occur only for $r>r_n$ (blue region). (b) Energy of formation of a bleb of detached radius $r$, Eq. \ref{eq energy}, for some values of the pressure $f$, including a typical equilibrium pressure $f_\textrm{eq}$ (solid lines). Membrane peeling may effectively strongly reduce the nucleation energy barrier (dashed lines). Parameter values are $\gamma=5\cdot 10^{-5}$ N/m \cite{Tinevez2009}, $d=30$ nm, $k=10^{-4}$ N/m \cite{Alert2015a}, $\delta=1$ nm \cite{Evans2001}, $\rho_0=10^{14}$ m$^{-2}$ \cite{Alert2015a}, $\chi=10^{-3}$ \cite{Rognoni2012}.}
\end{center}
\end{figure}

In turn, the unbound membrane is inflated by the intracellular pressure $f$ to become a spherical cap of radius $R_b=2\gamma/f$, as given by the Young-Laplace pressure drop \cite{Dai1999}. Then, the contact angle $\theta$ is geometrically related to the radius of the detached patch on the cortex, $r$, by $\sin\theta=r/R_b$ (see Fig. \ref{fig1}c). This implies that the vertical pulling of the membrane at the contact line reads $2\pi r\gamma\sin\theta=\pi r^2f$, namely the total force pushing on the unbound membrane, thus closing a relationship between $r$ and $u_c$ in Eq. \ref{eq force-balance}. Thereby, the critical stretching for peeling, $u^*$, translates into a critical size of the detached membrane region, $r_p$:
\begin{equation} \label{eq critical-radius}
r_p=2d\frac{f^*}{f},
\end{equation}
where $f^*=\rho_b^*ku^*$. Thus, $r_p$ is a critical radius for membrane peeling, and since the peeling process ends up in a mature bleb, this quantity indeed becomes a critical radius for bleb nucleation. Notably, $r_p$ is independent of membrane tension $\gamma$. Fig. \ref{fig2}a plots $r_p$ as a function of the pressure $f$ (red line), separating those detachments that grow to form a bleb by peeling (green and blue regions) from those that heal adhesion back (red region).

Remarkably, in contrast to classical nucleation, peeling is not controlled by the energy cost of forming a given nucleus, but instead by the kinetics of mem\-brane-cor\-tex linkers. This is apparent from the fact that the critical pressure $f^*$ is only a function of the kinetic parameter $\chi$, so that the critical radius in Eq. \ref{eq critical-radius} is completely determined by the kinetics of the linkers and the force $f$ they withstand. Indeed, the linkers at the contact line sustain the additional pulling due to the unbound membrane. Consequently, they may unbind even though the rest of the linkers remain below the detachment threshold, thereby unchaining the growth of the bleb. This effect was not captured by the classical nucleation approach to bleb formation \cite{Charras2008a,Norman2010,Sens2015,*Sheetz2006,*Lim2012}.

To compare our prediction to classical nucleation theory, we formulate the energy of bleb formation (see \cite{SM} for details):
\begin{equation} \label{eq energy}
E\left(r\right)\approx\pi r^2 w\left(f\right)-\frac{\pi f^2}{16\gamma}r^4,
\end{equation}
which is plotted in Fig. \ref{fig2}b for some values of the pressure $f$. Note that this energy includes the kinetics of the linkers via the pressure-dependent adhesion energy $w\left(f\right)$ introduced in Eq. 6 of Ref. \cite{Alert2015a}. Then, the maximum of the energy $E\left(r\right)$ indicates the classical nucleation radius, $r_n=\sqrt{8\gamma w\left(f\right)/f^2}$, which is also shown in Fig. \ref{fig2}a as a function of the pressure (blue line). This figure shows that membrane peeling may, for typical cellular parameters, require substantially smaller nucleation radii than classical energetic nucleation.

Finally, we stress that the classical mechanism is irrelevant even if the classical nucleation radius $r_n$ is smaller than $r_p$, since any radius $r$ such that $r_n<r<r_p$ would unavoidably shrink, even going uphill in energy. Similarly, for $r_p<r_n$, the growth of a bleb with $r_n>r>r_p$ also goes uphill in the energy landscape (see Fig. \ref{fig2}b). Therefore, bleb growth is not controlled by its global energy $E\left(r\right)$ but by the local dynamics of the contact line, and hence by linker kinetics. However, the probability of detaching a given patch by means of a fluctuation is still determined by the energy, Eq. \ref{eq energy}. Hence, bleb nucleation through membrane peeling entails overcoming an effective energy barrier $E\left(r_p\right)$, as shown in Fig. \ref{fig2}b (dashed lines). This effective barrier may typically be lower than the classical one \cite{SM}, thus strongly reducing the nucleation time.

\begin{figure*}[htbp]
\begin{center}
\includegraphics[width=\textwidth]{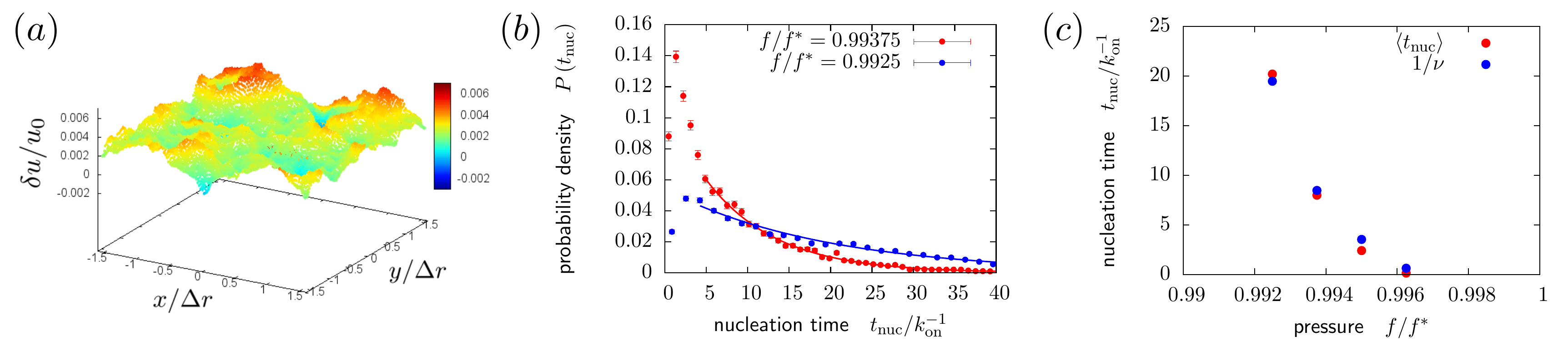}
\caption{\label{fig3}(Color online) Statistics of bleb nucleation times. (a) Snapshot of membrane undulations $\delta u$ from simulations. (b) Probability distribution of bleb nucleation times for two values of the pressure $f$. The long-time tails are fitted by an exponential $P\left(t_\textrm{nuc}\right)\sim e^{-\nu t_{\textrm{nuc}}}$. (c) The average bleb nucleation time decreases with pressure on the membrane, and so does the characteristic time scale of the process, $1/\nu$. Only pressures very close to the unbinding transition at $f^*$ are explored because of computational time limitations. In addition to those in Fig. \ref{fig2}, parameter values are $\kappa=10^{-19}$ J \cite{Dai1999}, $\eta_c=10^{-2}$ Pa$\cdot$s \cite{Charras2008a}, $\eta=50$ Pa$\cdot$s/$\mu$m \cite{Alert2015a}, $k_{\text{on}}=10^4$ s$^{-1}$ \cite{Rognoni2012}, $L=2$ $\mu$m, $n=1024$, $\Delta t=10^{-2}k_{\text{on}}$.}
\end{center}
\vskip-1.2cm
\end{figure*}

In the following, we formulate and simulate a continuum stochastic two-dimensional model of mem\-brane-cor\-tex adhesion, which will give access to the statistics of bleb nucleation times. The model describes the dynamics of membrane undulations at a linear level by coupling membrane elasticity and cytosol hydrodynamics to the force-dependent kinetics of mem\-brane-cor\-tex ligands \cite{Alert2015a}. Here, we add thermal and chemical fluctuations to trigger bleb nucleation. Hydrodynamic interactions render nonlocal dynamics for membrane undulations $\delta u\left(\vec{x}\right)$, which are decomposed in Fourier modes:
\begin{subequations}
\label{eq simulations}
\begin{multline}
\partial_t\delta\tilde{u}_{\vec{0}}=-\frac{1}{\eta}\left[\rho_b^{\textrm{eq}}k\delta\tilde{u}_{\vec{0}}+\frac{f}{\rho_b^{\textrm{eq}}}\delta\tilde{\rho}_{b,\vec{0}}\right]+\tilde{\zeta}_{\vec{0}}\left(t\right);\\
\left\langle\tilde{\zeta}_{\vec{0}}\left(t\right)\tilde{\zeta}_{\vec{0}}\left(t'\right)\right\rangle=\frac{2k_BT\pi\lambda^2}{\eta}\delta\left(t-t'\right), \label{eq sima}
\end{multline}
\begin{multline}
\partial_t\delta\tilde{u}_{\vec{q}}=\frac{-1}{4\eta_c q}\left[\left(\kappa q^4+\gamma q^2+\rho_b^{\textrm{eq}}k\right)\delta\tilde{u}_{\vec{q}}+\frac{f}{\rho_b^{\textrm{eq}}}\delta\tilde{\rho}_{b,\vec{q}}\right]+\\
+\tilde{\zeta}_{\vec{q}}\left(t\right);\quad \left\langle\tilde{\zeta}_{\vec{q}}\left(t\right)\tilde{\zeta}_{\vec{q}\,'}\left(t'\right)\right\rangle=\frac{k_BT}{2\eta_cq}\delta_{\vec{q},-\vec{q}\,'}\delta\left(t-t'\right), \label{eq simb}
\end{multline}
\end{subequations}
where $\vec{q}$ is the wave-vector, $\eta_c$ is the cytosol viscosity, $\kappa$ is the membrane bending rigidity, $\lambda$ is the correlation length of height fluctuations \cite{Alert2015a}, and $\rho_b^{\textrm{eq}}$ is the equilibrium density of bonds obtained from Eqs. \ref{eq 1}-\ref{eq kinetics}, with $f=\rho_b^{\textrm{eq}}ku_{\textrm{eq}}$. The dynamics of the $q=0$ mode, Eq. \ref{eq sima}, is decoupled from the rest, Eq. \ref{eq simb}, at the linear level. Eq. \ref{eq simb} includes thermal fluctuations in the form of a white noise, which is implemented in Fourier space \cite{Garcia-Ojalvo1992}. In turn, the kinetics of mem\-brane-cor\-tex linkers must include chemical fluctuations via a multiplicative noise term within the It\^{o} chemical Langevin equation approach \cite{Gillespie2000}, yielding
\begin{widetext}
\begin{multline} \label{eq multiplicative}
\partial_t\delta\rho_b\left(\vec{x}\right)=-\rho_b^{\textrm{eq}}\beta k\delta k_{\textrm{off}}^0e^{\beta ku_{\textrm{eq}}\delta}\delta u\left(\vec{x}\right)-\left[k_{\textrm{on}}+k_{\textrm{off}}^0 e^{\beta ku_{\textrm{eq}}\delta}\right]\delta\rho_b\left(\vec{x}\right)+\\
+\sqrt{k_{\textrm{on}}\left[\rho_0-\rho_b^{\textrm{eq}}\right]+\rho_b^{\textrm{eq}}k_{\textrm{off}}^0 e^{\beta ku_{\textrm{eq}}\delta}+\rho_b^{\textrm{eq}}\beta k\delta k_{\textrm{off}}^0 e^{\beta ku_{\textrm{eq}}\delta}\delta u\left(\vec{x}\right)+\left[k_{\textrm{off}}^0 e^{\beta ku_{\textrm{eq}}\delta}-k_{\textrm{on}}\right]\delta\rho_b\left(\vec{x}\right)}\;\frac{\Gamma\left(t\right)}{\sqrt{\pi\lambda_c^2}},
\end{multline}
\end{widetext}
with $\left\langle\Gamma\left(t\right)\Gamma\left(t'\right)\right\rangle=\delta\left(t-t'\right)$. 

Simulations of Eqs. \ref{eq simulations}-\ref{eq multiplicative} require two Fourier transforms at each time step to couple the dynamics of membrane undulations, which is evolved in Fourier space, to the kinetics of the linkers, evolved in real space. Therefore, our numerical procedure builds on the so-called Fourier space Brownian dynamics (FSBD) method \cite{Lin2004} for the simulation of continuum models of membrane dynamics. Simulations of a square membrane patch of side $L=n\Delta r=2\pi/\Delta q$ with periodic boundary conditions are performed. The membrane is initially fluctuating around the equilibrium position and link density corresponding to the chosen pressure $f$. A snapshot of the simulations is shown in Fig. \ref{fig3}a.

The simulation model is valid in the linear regime and hence cannot capture the complete formation of the bleb. However, it allows to determine the statistics of bleb nucleation, which reduces to the first-passage time statistics of finding a detached patch larger than the critical size. With this purpose, the following criterion is applied at each time step and around each point in space for which $\rho_b\left(\vec{x}_n,t\right)<\rho_b^*$. A bleb of detached radius $r$ is said to nucleate at point $\vec{x}_n$ at time $t$ if the average density of links within a circle of radius $r>r_p$ centered at $\vec{x}_n$ falls below the critical density for mem\-brane\--cor\-tex unbinding, $\left\langle\rho_b\left(\vec{x}-\vec{x}_n,t\right)\right\rangle_{\left|\vec{x}-\vec{x}_n\right|\leq r}<\rho_b^*$, while adhesion is restored within a slightly larger circle, $\left\langle\rho_b\left(\vec{x}-\vec{x}_n,t\right)\right\rangle_{\left|\vec{x}-\vec{x}_n\right|\leq r+\Delta r}>\rho_b^*$. Therefore, circles of increasing radius around the candidate points are considered until the nucleation criterion is fulfilled. A minimal radius $r_p\left(f\right)$ is demanded according to the critical nucleation radius in Eq. \ref{eq critical-radius} (red line in Fig. \ref{fig2}a). This nucleation criterion is fundamentally different from those in other simulation approaches to bleb formation, which either imposed an arbitrary maximal length of mem\-brane-cor\-tex linkers \cite{Spangler2011,*Taloni2015,Tyson2014} or removed some of them \cite{Young2010,*Strychalski2013,*Woolley2015}.

Employing our criterion, we have obtained the histogram of bleb nucleation times at a given pressure $f$, as exemplified in Fig. \ref{fig3}b. We find that the probability distribution of bleb nucleation times, $P\left(t_{\textrm{nuc}}\right)$, features an exponential tail $\sim e^{-\nu t_{\textrm{nuc}}}$ even for pressures very close to the unbinding transition $f^*$. This indicates that the process is dominated by a single time scale $1/\nu\propto e^{\beta E\left(r_p\right)}$ as usual in activation processes. Finally, Fig. \ref{fig3}c plots the decrease of the average nucleation time $\left\langle t_{\textrm{nuc}}\right\rangle$ with increasing pressure. The characteristic time $1/\nu$ obtained from the fits in Fig. \ref{fig3}b closely approaches the average $\left\langle t_{\textrm{nuc}}\right\rangle$, thus further stressing that it strongly dominates the kinetics of bleb nucleation.

Our results on the distribution of nucleation times for blebs (Fig. \ref{fig3}) are parallel to those reported for membrane adhesion in Figs. 4-5 of Ref. \cite{Bihr2012}.
In cells, linker aggregation or cortical remodelling are usually required to overcome the energy barrier associated to the nucleation of adhesion domains in reasonable time scales \cite{Zhang2008a,*Vink2013}.
In contrast, due to the reduced energy barrier essentially controlled by cortical tension (Fig. \ref{fig2}b), membrane peeling could allow bleb nucleation to proceed without them.

In summary, we 
have shown that membrane peeling governs bleb nucleation and can strongly enhance it. 
We have predicted the critical radius for bleb nucleation through membrane peeling, as well as its effective energy barrier, typically lower than that of classical nucleation theory. Our predictions can be experimentally tested by inducing local mem\-brane-cor\-tex detachments of controlled size, for instance via laser ablation of the cortex \cite{Tinevez2009} or via optogenetic control of either myosin activity or density of linkers. By means of simulations, we have also obtained the distribution of bleb nucleation times as a function of the stress on the membrane.
These results could also be assessed by measuring blebbing times in cells with perturbed cortical activity or subject to micropipette suction \cite{Merkel2000,Rentsch2000,Brugues2010,Sliogeryte2014,Tinevez2009}.

Our model for peeling sheds light on the mechanisms of homogeneous bleb nucleation, which may in general coexist with heterogeneous nucleation at preferential sites \cite{Tyson2014}. In future studies, our approach could be extended beyond the nucleation stage to study bleb growth and compare the results to experiments \cite{Charras2008a,Tinevez2009,Peukes2014} and simulations \cite{Young2010,Spangler2011,Strychalski2013,Tyson2014,Woolley2015,Taloni2015}. In addition, our simulation scheme could be employed to pursue the role of mem\-brane\--cor\-tex adhesion on the statistics of membrane fluctuations \cite{Peukes2014,Alert2015a}.
\begin{acknowledgments}
We thank P. Sens and J. Prost for illuminating discussions. R.A. acknowledges support from Fundaci\'{o} ``la Caixa''. We acknowledge support from MINECO under project FIS2013-41144-P and Generalitat de Catalunya under project 2014-SGR-878.
\end{acknowledgments}
\bibliography{Blebs}
\end{document}